\documentclass[namedreferences]{solarphysics}

\usepackage[hyperref,optionalrh,showbiblabels]{spr-sola-addons} 
\usepackage[optionalrh]{spr-sola-addons} 

\usepackage[hyperref,optionalrh,showbiblabels]{spr-sola-addons} 
\usepackage{graphicx}        

\usepackage{url}
\usepackage{hyperref}
\hypersetup{
    colorlinks=true,
    linkcolor=red,
    citecolor = blue,
    filecolor=cyan,
    urlcolor=magenta,
}  

\usepackage[normalem]{ulem} 

\usepackage{color}           
\usepackage{breakurl}        
\usepackage[hyperref,optionalrh,showbiblabels]{spr-sola-addons} 

\newcommand{\aap}{    {\it Astron. Astrophys.}}
\newcommand{\aaps}{   {\it Astron. Astrophys. Suppl.}}

\newcommand{\araa}{   {\it Ann. Rev. Astron. Astrophys.}}

\newcommand{\apj}{    {\it Astrophys. J.}}
\newcommand{\apjl}{   {\it Astrophys. J. Lett.}}
\newcommand{\apss}{   {\it Astrophys. Space Sci.}}

\newcommand{\mnras}{  {\it Mon. Not. Roy. Astron. Soc.}}

\newcommand{\solphys}{{\it Solar Phys.}}

\chardef\us=`\_
\begin{document}

\begin{article}

\begin{opening}

\title{Interferometric imaging, and beam-formed study of a moving Type IV Radio burst with LOFAR.}

\author[addressref={aff1,aff2}]{\inits{Hongyu}\fnm{Hongyu}~\lnm{Liu}}
\author[addressref=aff3]{\inits{Pietro}\fnm{Pietro}~\lnm{Zucca}}
\author[corref,addressref={aff1,aff4},email={kscho@kasi.re.kr}]{\inits{Kyung-suk}\fnm{Kyung-Suk}~\lnm{Cho}}
\author[addressref=aff5]{\fnm{Anshu}~\lnm{Kumari}}
\author[addressref={aff3,aff6}]{\fnm{Peijin}~\lnm{Zhang}}
\author[addressref={aff7,aff8}]{\fnm{Jasmina}~\lnm{Magdalenić}}
\author[addressref={aff1,aff4}]{\fnm{Rok-Soon}~\lnm{Kim}}
\author[addressref=aff1]{\fnm{Sujin}~\lnm{Kim}}
\author[addressref={aff1,aff9}]{\fnm{Juhyung}~\lnm{Kang}}
\address[id=aff1]{Korea Astronomy and Space Science Institute, 776 Daedeokdae-ro, Yuseong-gu, Daejeon 34055, Korea}
\address[id=aff2]{Center for Magnetic Materials and Devices, College of Physics and Electronic Engineering, Qujing Normal University, Qujing 655011, China}
\address[id=aff3]{ASTRON, the Netherlands Institute for Radio Astronomy, Oude Hoogeveensedijk 4, Dwingeloo 7991PD, the Netherlands}
\address[id=aff4]{University of Science and Technology, 217 Gajeong-ro, Yuseong-gu, Daejeon 34113, Korea}
\address[id=aff5]{Department of Physics, University of Helsinki, Pietari Kalmin katu 5, 00560 Helsinki, Finland}
\address[id=aff6]{Institute of Astronomy and National Astronomical Observatory, Bulgarian Academy of Sciences, Sofia 1784, Bulgaria}
\address[id=aff7]{Solar-Terrestrial Center of Excellence–SIDC, Royal Observatory of Belgium, Av. Circulaire 3, B-1180 Brussels, Belgium}

\address[id=aff8]{Center for mathematical Plasma Astrophysics, Department of Mathematics, KU Leuven, Celestijnenlaan 200B, B-3001 Leuven, Belgium}

\address[id=aff9]{Astronomy Program, Department of Physics and Astronomy, Seoul National University, Seoul 08826, Republic of Korea}


\runningauthor{H.\ Liu \emph{et al.}}
\runningtitle{Moving Type IV Radio burst with LOFAR observation}

\begin{abstract}
: Type IV radio burst has been studied for over 50 years. However, the specifics of the radio emission mechanisms is still an open question. In order to provide more information about the emission mechanisms, we studied a moving type IV radio burst with fine structures (spike group) by using the high resolution capability of Low-Frequency Array (LOFAR) on Aug 25, 2014 { (SOLA-D-21-00188)}. We present a comparison of Nan\c{c}ay RadioHeliograph (NRH) and the first LOFAR imaging data of type IV radio burst. The degree of circular polarization (DCP) is calculated at frequencies in the range 20$\sim$180 MHz using LOFAR data, and it was found that the value of DCP gradually increased during the event, with values of  10\%$\sim$20\%. LOFAR interferometric data were combined with white light observations in order to track the propagation of this type IV. The kinematics shows a westward motion of the radio sources, slower than the CME leading edge. The dynamic spectrum of LOFAR shows a large number of fine structures with duration of less than 1s and high brightness temperature ($T_\mathrm{B}$), i.e. $10^{12}$$\sim$$10^{13}$ K. The gradual increase of DCP supports gyrosynchrotron emission as the most plausible mechanism for the type IV. However, coherent emissions such as Electron Cyclotron Maser (ECM) instability can be responsible for small scale fine structures. Countless fine structures altogether were responsible for such high $T_\mathrm{B}$.
\end{abstract}
\keywords{Radio Bursts, Type IV; Radio Emission, Theory}
\end{opening}

\section{Introduction}
     \label{S-Introduction}
	 {The Sun is an active star, and its energetic phenomena} is frequently accompanied by electromagnetic emissions over a wide
spectral range including radio wavelengths.  {Solar radio
bursts can show up in dynamic spectrum (a plot of brightness temperature on a frequency-time plane) as the features of increased intensity
in comparison to the background emission.} They are classified at meter wavelengths into five major types, Type I to Type V named after Roman numbers \citep{1957CRAS..244.1326B}. Solar radio bursts are generally related to coronal mass ejections (CMEs) and solar flares \citep{2021A&A...647L..12M}.  {In this paper we focus on a type IV radio burst, which is a }long-lasting broadband continuum emission in radio wavelengths that usually appears approximately 10 minutes after solar flare \citep{1963IAUS...16..247P}. They are often considered to be generated by a process involving energetic particles trapped in a post flare loop  {or outward propagating magnetic structures, such as the CME magnetic fields \citep{1985srph.book..361S,Bastian2001,2019ApJ...870...30V,2021SoPh..296...38L}}. Depending on whether the radio source propagates, we can distinguish moving type IV radio burst (m-TypeIV) from stationary type IV radio burst \citep{1965ASSL....1..408K,1986SoPh..104...19P}. Due to its close relation with CMEs, m-TypeIV has been frequently addressed in solar radio physics since its discovery.

 {The emission }mechanism of type IV radio bursts is still highly controversial \cite[see e.g.][]{1974SoPh...36..157K,2019A&A...623A..63M,2021SoPh..296...38L}. \cite{1957CRAS..244.1326B} first observed type IV radio emission and proposed synchrotron emission as the mechanism. However, since high degree of circular polarization (DCP) ($>$20\%) is observed during some type IV events, \cite{1969PASA....1..189K} suggested gyrosynchrotron emission instead. \cite{1976ApJ...204..597B} concluded that type IV bursts of narrow bandwidth are generated by plasma emission instead of gyrosynchrotron emission. Later on, \cite{1981SoPh...73..191D} considered plasma emission to explain the observed high brightness temperature ($T_\mathrm{B}$). In addition to already mentioned emission mechanisms, \cite{1986ApJ...307..808W} discussed the possibility of electron cyclotron maser (ECM) emission { alongside the possibility of plasma emission (Upper Hybrid mode)}. Recent studies have featured gyrosynchrotron radiation \citep{2017A&A...608A.137C}, plasma emission \citep{2019ApJ...870...30V}, and ECM emission \citep{2018SoPh..293...58L} as a possible emission mechanisms of the type IV continuum.

One important characteristic of type IV continuum is its fine structures \citep{1987SoPh..112..347A,2011ASSL..375.....C}. Recent studies have reported fine structures of very short durations (4 to 60 ms at half power of the burst) \citep{2005ASSL..320..259M,2006ApJ...642L..77M,2008SoSyR..42..434C,2014ApJ...787...45K}. Fine structures may also have different appearances on radio dynamic spectrum. The most frequently observed fine structures of type IV continuum are zebra patterns \citep{2010Ap&SS.325..251T,2018ApJ...855L..29K}, fiber bursts \citep{1983ApJ...264..677B,2014RRPRA..19..295A}, spikes \citep{1986SoPh..104...99B}, and slowly drifting narrowband structures \citep{2014ApJ...787...45K}.  Since fine structures usually have higher brightness temperature than background type IV emission, they are more likely to be generated by a different emission process. 

Early studies of solar radio bursts were focused on understanding their basic physical characteristics employing single frequency observations and dynamic spectra. With the development of radio imaging instruments, more studies have been conducted to track the source of radio bursts on the Sun. Since the Nancay Radio-heliograph (NRH) started its observations in May, 1956 \citep{1957AnAp...20..155B}, metric-wavelength (i.e. $<$500MHz) imaging study of the radio Sun has a history of more than 60 years. Culgoora radio-heliograph was the second one of the type, starting observations from Feb 1968 \citep{1970PASA....1..365W}, but imaging observations stopped working in 1980s. In 1997, Gauribidanur Radio-heliograph \cite[GRAPH;][]{1998SoPh..181..439R} in India came into use, as an addition to the existing NRH instrument that  can observe Stokes I and V. In January, 2015, NRH imaging observations stopped, to undergo extensive maintenance. Observations were resumed in November 2020, and currently only Stokes I is observed, as well as the radio image of the Sun at ten separate frequencies within the range of  150 MHz-445 MHz \citep{1997LNP...483..192K}. With the help of NRH,  {a number of} studies have been carried out addressing type IV continuum
emission \citep{2016A&A...586A..29B,2016ApJ...826..125K}. The Low-Frequency Array (LOFAR) \citep{2013A&A...556A...2V} in the Netherlands is not a solar dedicated instrument but during regular solar dedicated campaigns, the Sun  is observed in a similar frequency range as by NRH. There have been several studies on solar type III \citep[e.g., ][]{2014A&A...568A..67M,2018A&A...614A..69R, 2020A&A...639A.115Z, morosan2022}, type II radio bursts \citep[e.g., ][]{2018ApJ...868...79C,2018A&A...615A..89Z,2020ApJ...897L..15M,2021ApJ...909....2M}, and short duration bursts \citep[e.g., ][]{2015A&A...580A..65M,2020ApJ...891...89Z} using LOFAR data. Type IV radio bursts on the other hand, have been reported less often with LOFAR observations. \cite{2019ApJ...873...48G} performed a statistical study of { the positions} of solar radio bursts, among which two were type IV events. The authors found a discrepancy between the obtained radio source locations and the expected radial height as mapped by the Newkirk density model. This discrepancy was attributed to the strong wave scattering due to plasma turbulence in the active corona.

In our study, we  {examine} a moving type IV radio burst with its fine structures observed by LOFAR on August 25, 2014.
Despite of its long history, the emission mechanism of Type IV bursts is not fully understood.  {This study will contribute to a better understanding of type IV bursts based on the high resolution observation of LOFAR.}
In Section \ref{S2}, we showed an overview of the type IV radio burst and associated solar activity. A comparison of NRH and LOFAR imaging data is then presented.

In Section \ref{S3}, we calculated the degree of circular polarization of the type IV. In addition, we combined the LOFAR imaging data with SDO-AIA and Solar and Heliospheric Observatory/Large Angle and Spectrometric Coronagraph (SOHO/LASCO)-C2 to track the type IV radio sources and understand their association with the ambient coronal structures. Finally, we presented details of fine structures and the estimated brightness temperature, followed by discussion on the radiation mechanism. This is the first detailed analysis of a type IV radio burst with LOFAR interferometric imaging data.

\section{Observations and Data Analysis}
      \label{S2}

	\subsection{   The Low-Frequency Array and Data Preparation}
\label{S21}
 {Low-Frequency Array (LOFAR) utilizes omni-directional antennas to form a phased array}. LOFAR operates in the 10 MHz to 240 MHz frequency range with a frequency resolution of 12.5kHz. There are two types of antennas: Low Band Antenna (LBA) and High Band Antenna (HBA), observing two different frequency ranges 10-90 MHz and 110-240 MHz, respectively. The frequency range is subject to change according to the LOFAR stations used to carry out the observation. While conventional radioheliographs like NRH can only provide Stokes I and V { data}, LOFAR simultaneously observes Stokes I, Q, U and V \citep{2013A&A...556A...2V}. As one of the newest radio telescopes to date, LOFAR greatly improved the angular resolutions of solar observations \citep{2011pre7.conf..507M}.

We carried out imaging spectroscopy observations of the Sun with  LOFAR on 25 August 2014. For LOFAR observations of this event, the frequency range of LBA is 10 MHz to 90 MHz, and that of HBA is 110 MHz to 190 MHz.  
The dynamic spectrum in this work is obtained from core station. The flux value in the dynamic spectrum is un-calibrated relative flux intensity with median background subtracted for each frequency channel, the combined dynamic spectrum of LBA and HBA is shown in Figure \ref{fig1}.

The interferometric imaging used the data from both LOFAR core and remote stations,  imaging data is available only for HBA in this event.
The measurement set is pre-processed with the Default Pre-Processing Pipeline (DPPP; \citealp{2018ascl.soft04003V}). The  pre-process includes three steps, to:  (1) derive the amplitude and the phase solutions from the analysis of the observation of Virgo-A with a Virgo-A model;  {(2) apply the solutions to the solar observations to obtain the calibrated visibilities;} (3) apply the LOFAR-beam corrections to the calibrated visibility. After the pre-processing, we used the WS-Clean \citep{2014MNRAS.444..606O,2017MNRAS.471..301O} for the Fourier transform and deconvolution to obtain the radio flux image. 
The calibration in SFU is a straightforward calculation inferred by knowing the beam size of LOFAR and by observing at the same time a calibrator.
For more details, see e.g. \citealp{2009IEEEP..97.1431D}.
 {The imaging and calibration process to obtain the physical unit (brightness temperature in [K]) uses the LOFAR-Sun toolkit with similar procedures as recent solar radio imaging works \citep{2020A&A...639A.115Z,2022ApJ..932.17} .}

The coordinate system of the imaging result is transformed from Equinox with J2000 epoch to helioprojective coordinate system.
On the lower-right corner in the radio image of this day, there are some projected fake sources. So we have also applied a fade-out fit to the calibrated data on the lower-right corner to erase the fake sources. This doesn't affect the actual radio sources of the event. All the processes mentioned above were done with the ASTRON CEP3 server, and visualization was done using the Solar and Space Weather KSP routines (https://git.astron.nl/ssw-ksp/lofar-sun-tools) using SunPy codes \citep{2015CS&D....8a4009S}.

\subsection{Event overview}
\label{S22}

\begin{figure}
	\centering
	\includegraphics[trim=0.2cm 0cm 0cm 0cm, clip=true, width=12cm, angle=0]{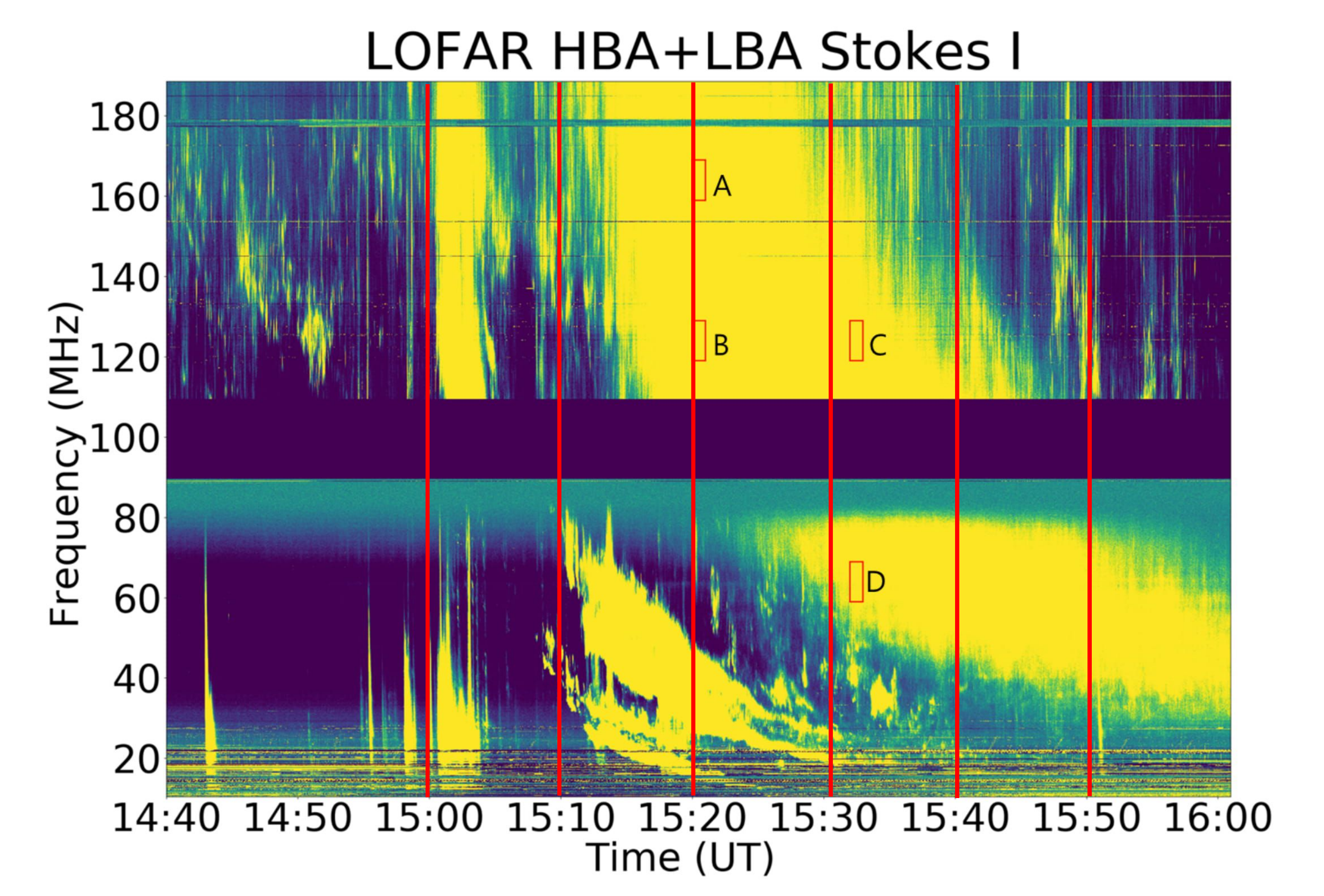}
	\caption{LOFAR radio dynamic spectrum on Aug 25, 2014 from 14:40 UT to 16:00 UT. Type III Radio bursts appeared in both HBA (110-190 MHz) and LBA (10-90 MHz) at 15:00 UT, then a type II radio burst appeared in LBA at 15:09 UT. At 15:16, a type IV radio burst appeared in HBA and later in LBA. A, B, C and D are 4 selected areas within the type IV radio burst, and zoom-in plots of these areas are shown in Figure \ref{fig6}. The vertical lines on the spectrum in red color are the time stamps where the polarization profiles were estimated in Figure \ref{fig35}.}
	\label{fig1}
\end{figure}
On Aug 25, 2014, an
M2.0 class flare  {occurred in NOAA 12146}, starting at 14:46 UT\footnote{\url{https://www.solarmonitor.org/index.php?date=20140825&region=12146}} (SOL2014-08-25T15:10). Meanwhile, SOHO/LASCO CME catalog provided by NASA Goddard Space Flight Center \citep{2009EM&P..104..295G}, reports a halo CME first seen in the  SOHO/LASCO C2 field of view at 15:36 UT\footnote{\url{https://cdaw.gsfc.nasa.gov/movie/make_javamovie.php?stime=20140825_1401&etime=20140825_1907&img1=lasc2rdf&title=20140825.153605.p270g;V=555km/s}}. The CME speed is reported to be 555 km/s. The eruptive CME/flare event was  {associated with a complex type III/II/IV radio burst shown} by LOFAR combined dynamic spectrum (HBA and LBA) in Figure \ref{fig1}.

In Figure \ref{fig1}, a group of type III radio bursts associated with the impulsive phase of the long duration flare is visible on both HBA and LBA at 15:00 UT, followed by a type II radio burst on LBA at 15:09 UT \citep{2020ApJ...897L..15M}. At 15:10 UT, a type IV radio burst was observed, first in HBA and later in LBA. Despite the observation gap between 90 MHz and 110 MHz, it is clear that the type IV radio continuum spans from the HBA to the LBA range. The frequency drift of the continuum emission implies that this is a moving type IV radio burst. Another group of type III bursts was observed at about 15:48 UT in HBA. In this study, we focus only on the type IV emission.

\subsection{LOFAR imaging data of the Sun and comparison with NRH}
\label{S23}

\begin{figure}
	\centering
	\includegraphics[trim=0cm 0cm 0cm 0cm, clip=true, width=12cm, angle=0]{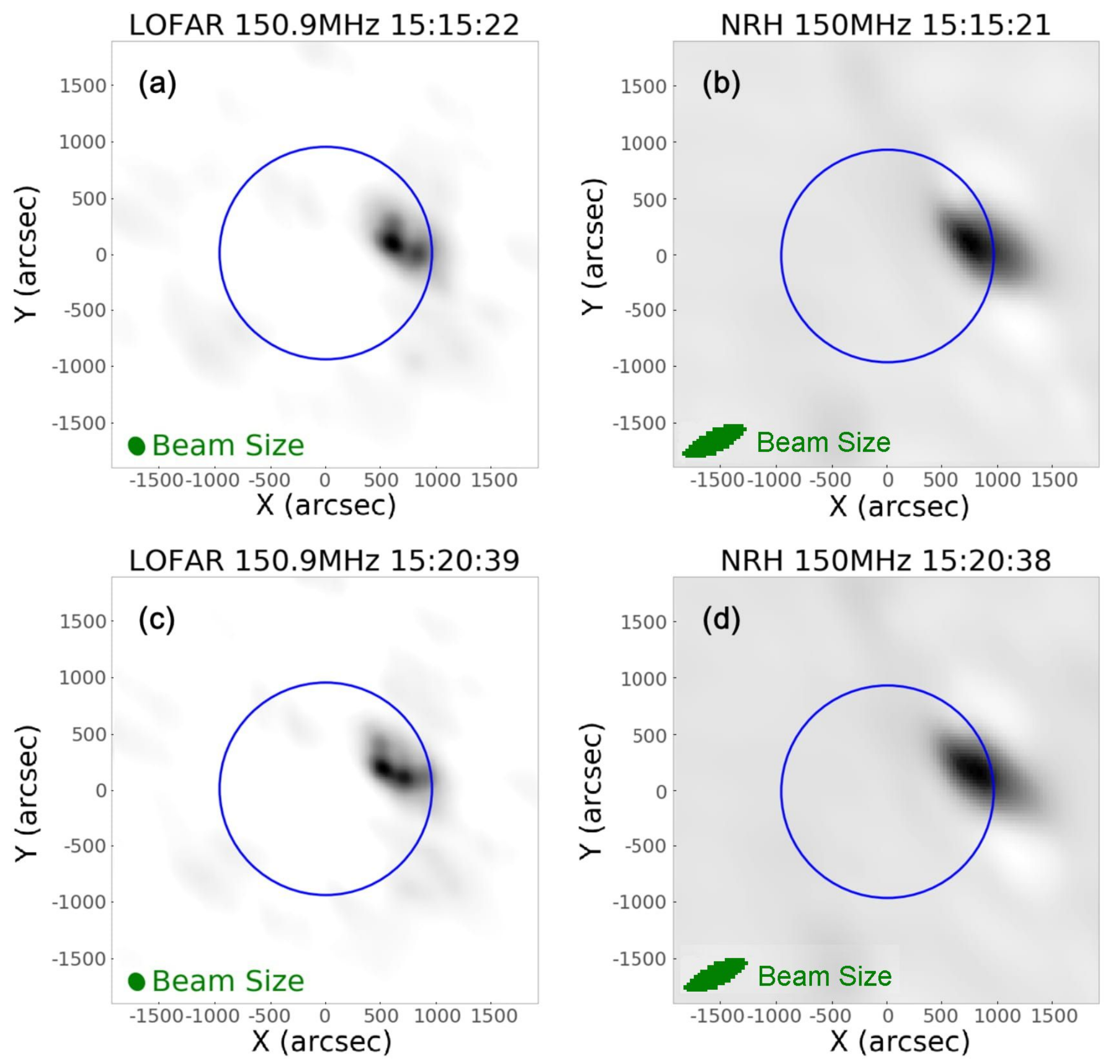}
	\caption{Comparison of LOFAR and NRH imaging observations of the Sun at 150.9 MHz on Aug 25, 2014. (a) and (b) are respectively LOFAR and NRH observations of the Sun at 15:15 UT while (c) and (d) are the same observations at 15:20 UT. The solar limb is indicated by a blue circle. Beam sizes of LOFAR and NRH are indicated by green ellipses in (a) and (c), and (b) and (d), respectively. All images represent the position of the type IV continuum source. (An animation of this figure is available online.)
	}
	\label{fig2}
\end{figure}

 {We performed a detailed comparison of LOFAR and NRH imaging data.
The comparison of NRH and LOFAR spatial resolution is important because we employ LOFAR observation with the remote baselines to demonstrate the advantages in beam size and resolution of a long baseline.
}
For this event, LOFAR HBA has observations from 110 to 190 MHz. Two NRH frequency channels (150.9 MHz and 173 MHz), overlap with LOFAR observations. On Aug 25, 2014, LOFAR had data from 12:12 UT until 16:12 UT, while NRH observed the Sun until 15:23 UT. Since LOFAR has a frequency resolution of 12.5 kHz, radio imaging data can be produced to match the exact NRH frequency.
After acquiring the cleaned LOFAR imaging data of the Sun, we re-scaled LOFAR data to make the field of view identical to that of NRH, and plotted LOFAR 150 MHz and NRH 150 MHz observations together. A movie is available in the online material. Some of the frames are shown in Figure \ref{fig2}.

Both movie and images show that the positions of LOFAR and NRH radio sources are in agreement. In Figure \ref{fig2} (a), we can see multiple separate sources on LOFAR while in  Figure \ref{fig2} (b), we can only see one faint source on NRH. Similarly, in Figure \ref{fig2} (c), at least 3 separate structures are visible on LOFAR images, while in Figure \ref{fig2} (d), we can only see a blurry radio source. The beam size ($\sim$ 200 arcsec along major axes) of LOFAR is shown on the lower-left corner in Figure \ref{fig2} (a) and (c),
while beam size ($\sim$ 650 arcsec along major axes) of NRH is shown on the lower-left corner in Figure \ref{fig2} (b) and (d). Since the beam size of LOFAR is smaller than the displacement of the separate radio sources ($\sim$ 500 arcsec), we clearly observe two sources that are separated and resolved.
We are able to resolve the multiple sources at the onset of the Type IV which are appearing as a single large source in NRH with ~1.8 km baseline.
This indicates the importance of medium baselines 3 to 10 Km to resolve complex radio sources and understand better the evolution and kinematics of the event.

\section{Results}
\label{S3}

\subsection{Degree of circular polarization}
\label{S31}

\begin{figure}
	\centering
	\includegraphics[trim=0.2cm 0cm 0cm 0cm, clip=true, width=11cm, angle=0]{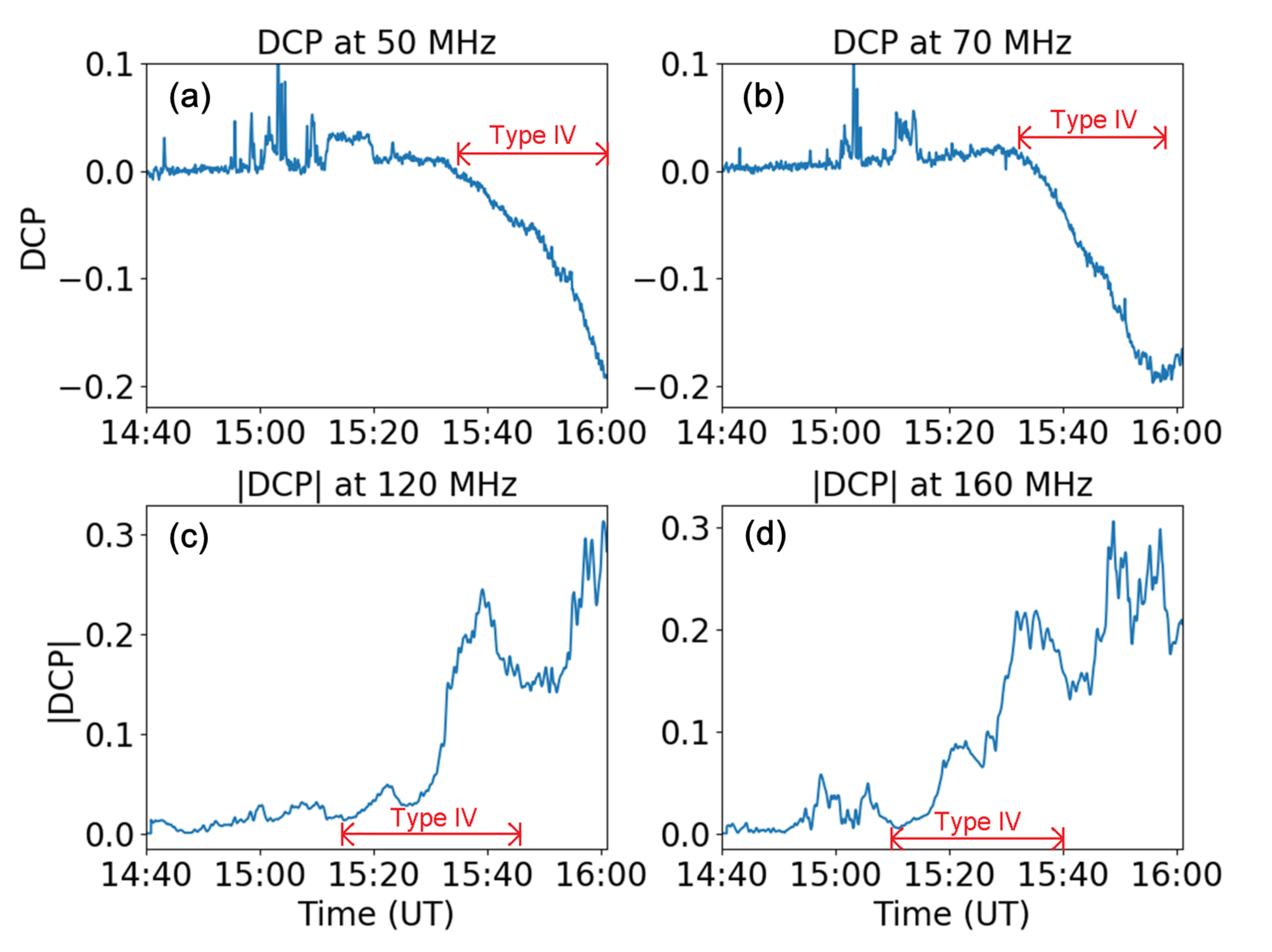}
	\caption{Degree of circular polarization at 4 frequencies. (a) and (b) are DCP at 50 MHz and 70 MHz respectively; (c) and (d) are the absolute values of DCPs at 120 MHz and 160 MHz. The time range is the same with that of Figure \ref{fig1}. Time span of type IV is shown on each panel.  {Note that the difference in the type IV span between (a), (b) and (c), (d) is due to the frequency drift of the burst.}}
	\label{fig3}
\end{figure}
A solar radio burst has many physical parameters, and they are crucial not only to determine the emission mechanism, but also to understand the ambient coronal conditions \citep{1975SoPh...43..211M}. One of the important parameters of the radio bursts is the degree of circular polarization (DCP). The degree of circular polarization of solar emission is given by \citep{1992plfa.book.....C}:
\begin{equation}
\label{eq1}
DCP  = \frac{V}{I} ,
\end{equation}
where V and I are the corresponding Stokes parameters. LOFAR provides full Stokes parameters, i.e. Stokes I, Q, U, and V.
As often observed for radio instrumentation, a small amount of Stokes V on the LOFAR HBA band has leaked into Stokes U.
In fact, the polarization leakage (linear spill of intensity from Stoke V to U) is a common instrumental issue in the majority of radio instruments.
We simply correct the leakage by using the following process since LOFAR provide four Stokes parameters.
In order to obtain the real Stokes V data in HBA, we introduced some corrections. Thus, the DCP of HBA is calculated as follows:
\begin{equation}
\label{eq2}
\vert DCP_\mathrm{corr}\vert  = \frac{\sqrt{V^2+U^2}}{I} .
\end{equation}

The signal leakage to Stokes U is proportional to the Stokes V intensity, making it a clear instrumental effect.
Moreover, this correction can be confidently done for the Sun as we can assume
that linear component of the incoming radio waves are removed by Faraday rotation (FR) in the corona.
Figure \ref{fig3} shows DCP diagrams from 14:40 UT to 16:00 UT at 50 MHz and 70 MHz calculated using Eq. (\ref{eq1}).
The presented absolute value of DCP at 120 MHz and 160 MHz was estimated using Eq. (\ref{eq2}).

Figure \ref{fig3} shows that the DCP of this type IV continuum  {amounts 10\%$\sim$20\%}, and the DCP value rises as type IV develops in time.
The increasing DCP after the end of the Type IV burst as shown in Figure 3 (c) and (d),
is due to the presence of other types of radio bursts superposed at higher frequencies.
The Type I and Type III like sources result in the increase of DCP visible.
This is a good indicator that the instrument is capable of measuring DCP well.
The increase at 15:40 UT at 120 MHz is a clear sign of the superposition of the type III and Type I sources.
We can still show well that the type IV (especially where well isolated in the profiles at 50 and 70 MHz) show a clear trend of decreasing DCP.
It can be done thanks to the full Stokes observations of LOFAR.
From \ref{fig3} (a) and (b), we conclude that the DCP of this type IV is negative, which means that left hand circular polarization dominates.
To preserve the sign of the DCP we use the following equation:

\begin{equation}
\label{eq3}
DCP_\mathrm{corr}  =\frac{\vert V\vert}{V}  \frac{\sqrt{V^2+U^2}}{I} .
\end{equation}

We also plot the DCP along vertical lines on the spectrum in Figure \ref{fig1}.
For this, we have selected 6 time instances from the spectra, which includes the DCP
during no bursts observed and during the type II, III and IV events.
The DCP is calculated by averaging the 1 min spectral data and implementing the correction as shown in Eq. (3)

\begin{figure*}
	\centering
	\includegraphics[trim=0cm 0cm 0cm 0cm, clip=true, width=1\linewidth, angle=0]{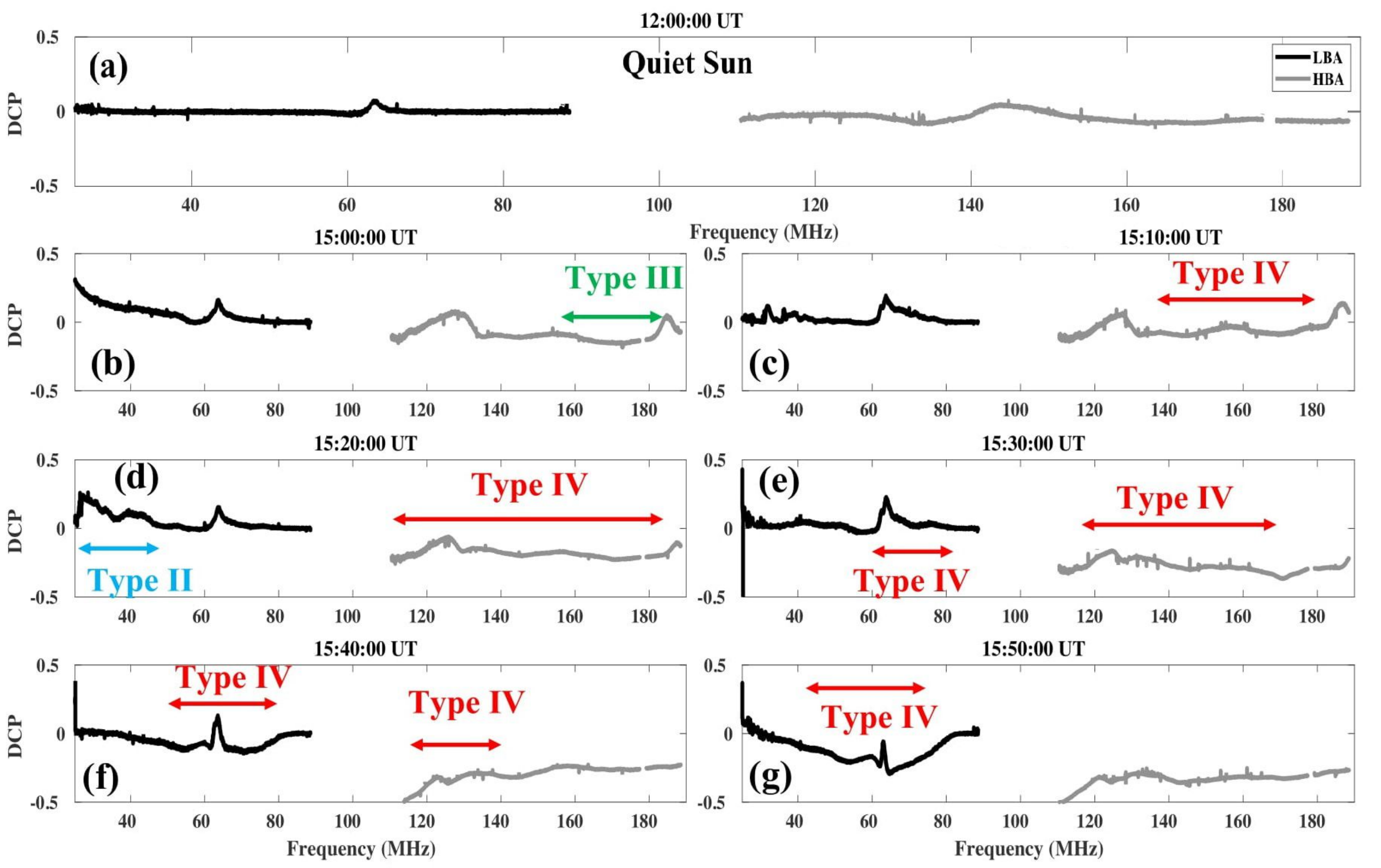}	
	\caption{(a): DCP variation along the full LOFAR frequency range at 12:00 UT where no bursts are observed (Quiet Sun); (b): Corrected DCP along the full LOFAR frequency range during type III radio bursts in HBA; (c) to (g): Corrected DCP along the full LOFAR frequency range during type IV event, for five different instances; and (d): Corrected DCP during type II event in LBA.}
	\label{fig35}
\end{figure*}

Figure \ref{fig35} (a) shows the DCP variation along the full LOFAR frequency range at 12:00 UT when no bursts are present.
The flat profile with very little of modulation shows that in the absence of radio burst, the instrument records almost no DCP.
Panels (b) to (g) of Figure \ref{fig35} show the DCPs (again along the full LOFAR frequency range) at six different times when there are radio bursts recorded.
The time intervals during which different types of radio bursts were recorded are marked with arrows of different colors.
Starting from about 15:00 UT a group of type III radio bursts is visible on the spectrum, while from 15:10 UT to 15:50 UT,
the type IV which we study is present (see also Fig.1). A gradual change of the DCP towards a negative value is observed
as the type IV radio burst propagates away from the Sun (Figure \ref{fig35} panels (d) to (g)).
We note that at 63 MHz, there is a spike which can be seen at all considered times. This spike is due to radio frequency interference (RFI).

Although the observed gradual rise of DCP value favors gyrosynchrotron radiation as the generation mechanism \citep{1985ARA&A..23..169D},
other physical parameters such as $T_\mathrm{B}$, also need to be considered in order to determine the emission mechanism of this type IV continuum.

\subsection{Propagation of radio sources}
\label{S32}

\begin{figure*}
	\centering
	\includegraphics[trim=0cm 0cm 0cm 0cm, clip=true, width=12.2cm, angle=0]{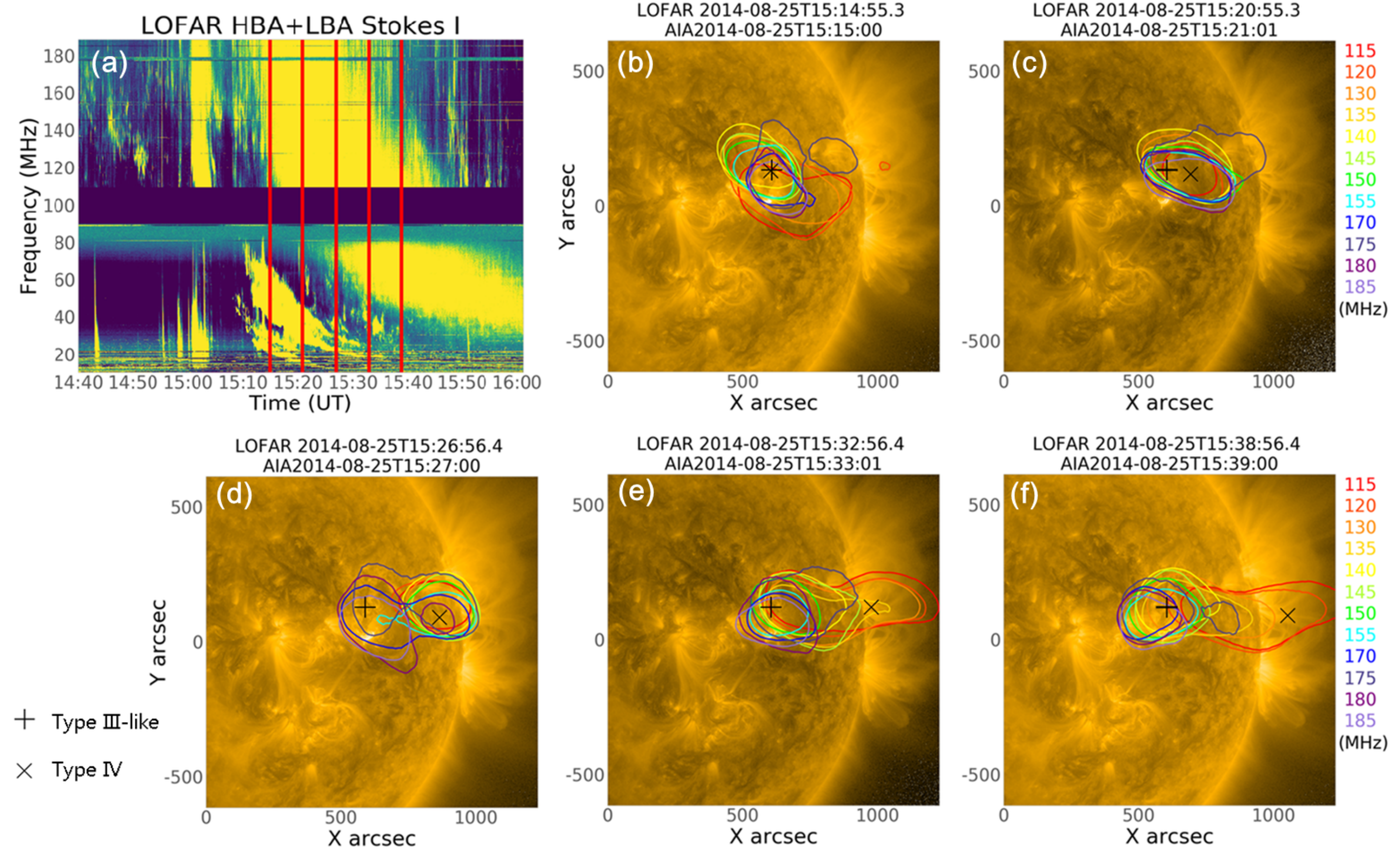}
	\caption{(a): LOFAR HBA and LBA dynamic spectrum on Aug 25, 2014. Five red vertical lines indicate the time of  panels (b)--(f). On the panels (b)--(f): SDO-AIA 171\AA~ images are over-plotted with LOFAR 70\% maximum contours at every 5 MHz from 115 to 185 MHz, with the exception of 125, 160 and 165 MHz. (b): All LOFAR frequencies start to group near the active region; (c): Frequency contours propagate to the west limb of the Sun; (d): High frequency (170$\sim$185 MHz) contours start to warp back to the active region, from where Type III-like structures originated; (e) and (f): only 3 low frequency contours keep propagating westward while all remaining frequencies are again near the active region. (An animation of this figure is available online.)
	}
	\label{fig4}
\end{figure*}
\begin{figure*}
	\centering
	\includegraphics[trim=0.19cm 0.1cm 0.19cm 0.1cm, clip=true, width=12.2cm, angle=0]{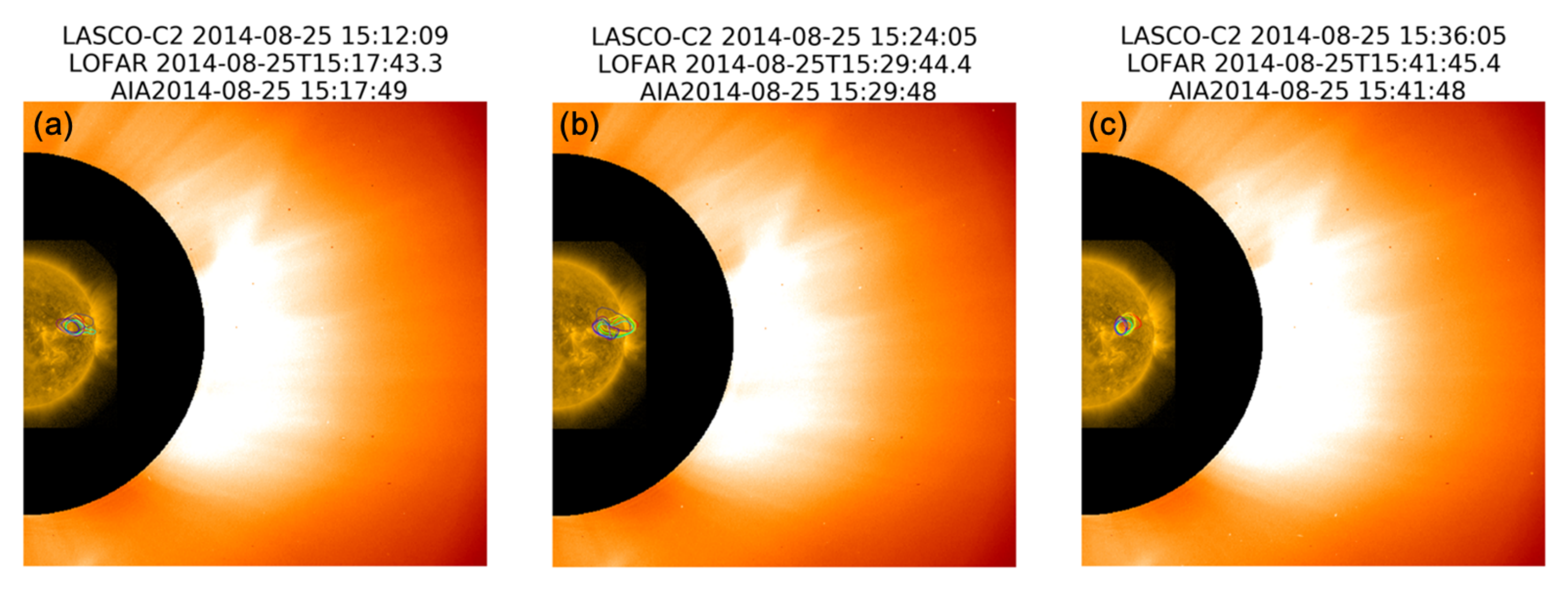}
	\caption{Combined plot of LASCO-C2, LOFAR radio source and SDO-AIA 171 \AA~. A CME is visible on LASCO-C2 in the direction of the propagation of the type IV.}
	\label{fig5}
\end{figure*}

In order to investigate the properties of the type IV continuum and its association with the ambient coronal structures, we plotted LOFAR 70\% maximum contours of different frequencies on SDO-AIA 171 \AA~ images. The movie is provided as the online material and few frames are shown in Figure \ref{fig4} (b) $\sim$ (f). The LOFAR channels at 125, 160 and 165 MHz suffer from the calibration issues, { and we couldn't identify the radio sources, }so we have excluded them from the study.

At 15:15 UT, all radio sources appeared to be situated above the active region (NOAA AR 12146) that was the source of the associated CME/flare event. In Figure \ref{fig4} (c), almost all frequencies seems to propagate toward the west limb of the Sun. At 15:27 UT, radio sources at higher frequencies (170$\sim$185 MHz) started to be observed again closer to the active region. This behavior of the radio sources is expected because the continuum emission decreases in intensity and is no longer observed at higher LOFAR frequencies (see Figure \ref{fig1}). We therefore, start to observe again the radio sources associated with the active region. At the same time, type IV emission at lower frequencies of the HBA band can still be observed until about 15:48~UT. In Figure \ref{fig4} (e) and (f), only 3 low frequency (115$\sim$130 MHz) contours kept propagating westward while all remaining frequencies map the faint radio source, associated with the active region. An expanding post-flare loop was not clearly seen in AIA 171\AA~ running differential images. It might be due to the low emissivity of such loops in EUV wavelengths.

Figure \ref{fig5} shows this event in a larger scale, with LOFAR, AIA and LASCO-C2 on the same panel. We can notice the CME eruption in the direction of the propagation of type IV radio source.
This CME eruption continued after the type IV disappeared in LOFAR HBA frequency range. Unfortunately, LBA radio imaging was not active for this observation and we cannot confirm the association of the lower frequencies radio sources of the type IV continuum with the CME core.

\subsection{Fine structures and emission mechanism}
\label{S33}

\begin{figure*}
	\centering
	\includegraphics[trim=0cm 0cm 0cm 0cm, clip=true, width=12cm, angle=0]{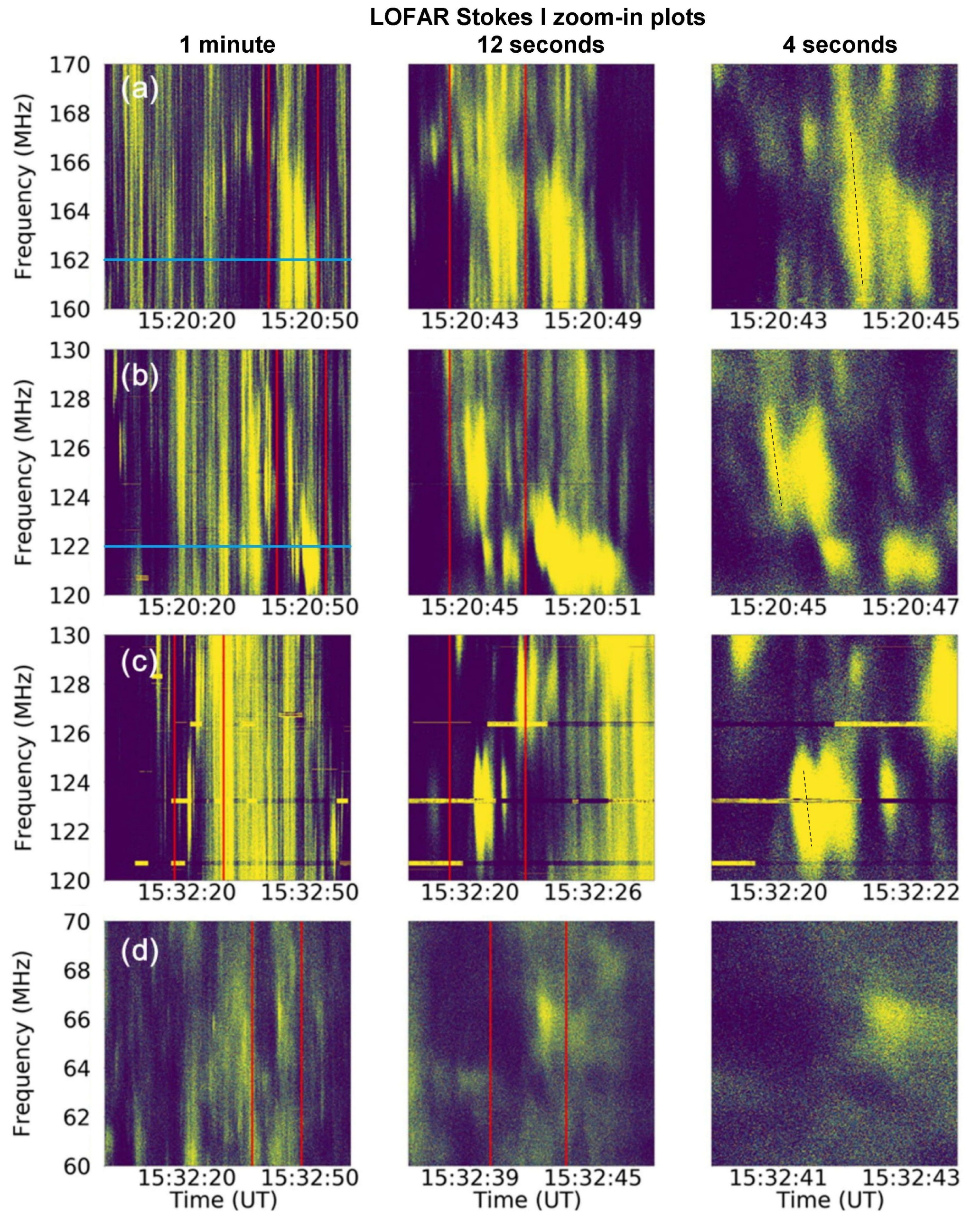}
	\caption{	 {Details of regions A-D of the dynamic spectrum in Fig. 1.} Spectra on left columns of (a), (b), (c) and (d) are 1 minute zoom-in plots of the areas depicted by red box A, B, C and D in Figure \ref{fig1}; Middle columns of (a), (b), (c) and (d) are 12 seconds zoom-in plots of areas between two red lines on left columns of (a), (b), (c) and (d) respectively; Right columns are 4 seconds zoom-in plots of areas between the red lines on middle columns. }
	\label{fig6}
\end{figure*}

\begin{figure}
	\centering
	\includegraphics[trim=0.1cm 0cm 0.38cm 0cm, clip=true, width=9.1cm, angle=0]{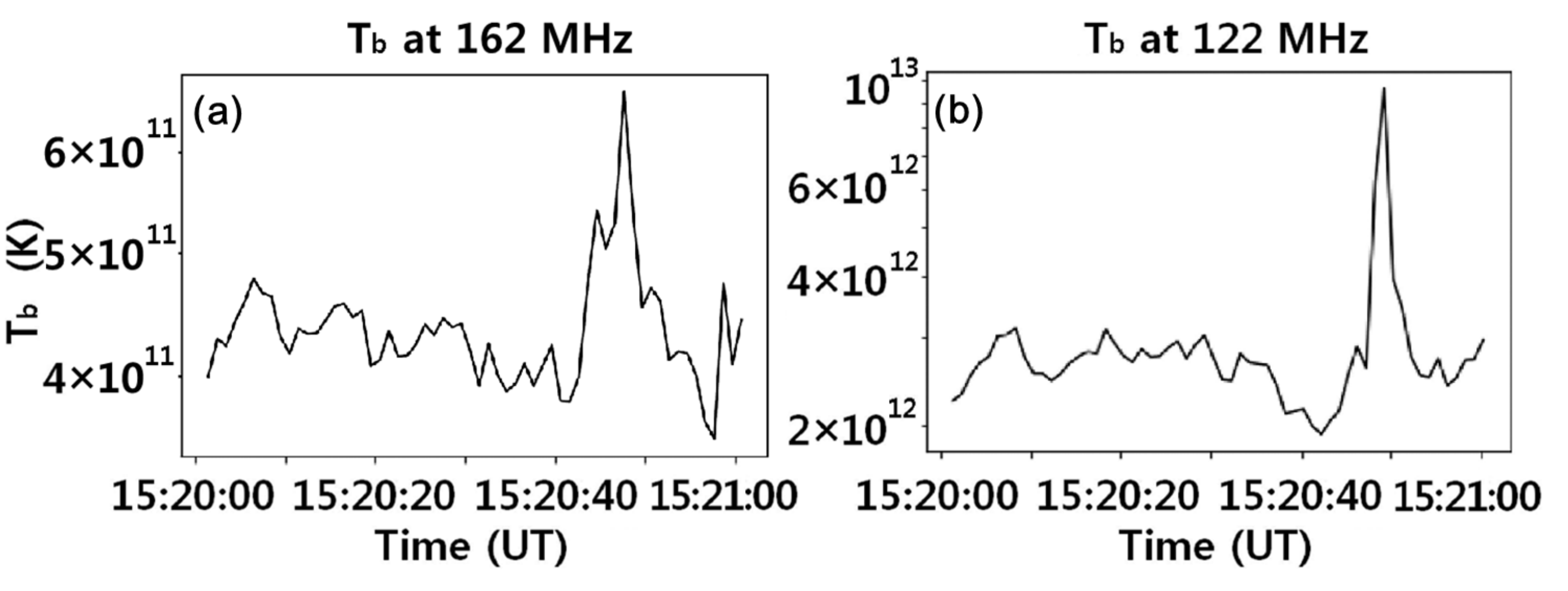}
	\caption{
		Brightness temperature profiles at 162 MHz and 122 MHz along blue lines in left panels of Figure \ref{fig6} (a) and (b) respectively.}
	\label{fig7}
\end{figure}

 {As already mentioned in Section 1}, a type IV continuum can have superposed fine structures which have duration much shorter than the type IV itself.
We analyzed the type IV fine structure in order to obtain more knowledge on the possible origin. We observe different fine structures, such as spikes, drifting spike and patchy, irregular fine structures.
To isolate short duration fine structures, we have plotted the dynamic spectrum in a shorter time range for selected areas within this type IV radio burst.

Figure \ref{fig6} (a), (b), (c) and (d)  {are details} of the dynamic spectrum within time and frequency ranges indicated by red boxes A, B, C and D in Figure \ref{fig1}.
Time ranges of left, middle and right columns are 1 minute, 12 seconds and 4 seconds respectively.
Despite the short time range, we can see plenty of fine structures that last for about 1s and less.
These narrow-band bursts have the characteristics of spikes, some of them show positive frequency drifts, while others show negative frequency drifts \citep{2016A&A...586A..29B}. 
We measure the drift rate of the fine structure as denoted with a dotted line in the right columns of Figure 7.
We find that the mean draft rate is about 20 MHz/s.

Brightness temperature plays an important role in determining the emission mechanism. It can be obtained according to Rayleigh-Jeans law:
\begin{equation}
\label{eq3}
T_\mathrm{B}  = \frac{\lambda^2}{2k\Omega}S ,
\end{equation}
where $T_\mathrm{B}$ is the brightness temperature, $\lambda$ is the wavelength, k is the Boltzmann constant, $\Omega$ is the beam solid angle, and S is the flux density. We plotted the $T_\mathrm{B}$ profiles at 162 MHz and 122 MHz from 15:20 UT to 15:21 UT in Figure \ref{fig7} (a) and (b). $T_\mathrm{B}$ data provided by LOFAR have a time resolution of 1 s. Note that these plots are along 2 blue parallel lines in left panels of Figure \ref{fig6} (a) and (b). As is shown in Figure \ref{fig6}, there are plenty of fine structures that are brighter than the background type IV. However, only the most intense fine structures have the brightest temperature noticeably different with the type IV continuum. The intense bursts at 162 MHz reach about $6.5\times10^{11}$ K, while the brightness temperature of the type IV background and the weaker fine structures fluctuates around $4\times10^{11}$ K. Situation is similar at 122 MHz but, with in general higher brightness temperatures. The intense burst has brightness temperature of about $10^{13}$ K, while background and weaker fine structures have value of about $2.5\times10^{12}$ K. These examples show that in a case of faint fine structures the brightness temperatures will be on a similar  {order} as for the type IV continuum.

	\section{Discussion and Conclusions}
\label{S4}
A moving type IV radio burst was observed at LOFAR dynamic spectrum on August 25, 2014.
We performed a detailed comparison of NRH and LOFAR imaging.
Having a higher spatial resolution LOFAR observed two separate sources at the onset of the m-TypeIV, compared to one unique larger source on NRH.
Using the full stokes parameters from the LOFAR dynamic spectra, we calculated the degree of circular polarization during the propagation of this moving type IV.
The DCP of this type IV lies in the range 10\%$\sim$20\% and it is left hand polarized. The DCP value showed a gradual increase with the development of the type IV.
In addition, the combined LOFAR interferometric data with SDO-AIA 171 and LASCO-C2 indicate that this type IV is generated on expanding flare loops located in the CME core.
 {The close-in details} of the LOFAR dynamic spectrum showed the existence of countless fine structures.
Majority of them look like narrow-band drifting spikes and broadband pulses, and they generally last for less than one second.
$T_\mathrm{B}$ analysis showed that the most intense fine structures can exhibit a $T_\mathrm{B}$ as high as $10^{13}$ K.
The type IV continuum also showed higher than expected $T_\mathrm{B}$ (up to $10^{12}$ K).

We can consider plasma emission, gyrosynchrotron, and ECM as generation mechanisms of type IV background and the fine structure.
Our observation presented extremely high brightness temperature and low polarization degree with gradual increase.
Regarding type IV background, the observed gradual rise of DCP value favors gyrosynchrotron radiation as the generation mechanism \citep{1985ARA&A..23..169D} because the gradual increase of DCP as shown in Figure \ref{fig3} can't be explained by original plasma emission models \citep{1985srph.book..361S}. One possible argument against the gyrosynchrotron emission is that the emission mechanism has difficulty in explaining high $T_\mathrm{B}$. { \cite{1978SoPh...60..383R} considered the local electron density and claimed that $T_\mathrm{B}$ of gyrosynchrotron emission can't exceed $10^{9}$ K if the source is moderately polarized ($>$40$\%$)}. We note that DCP is not very high (10\%$\sim$20\%), and if the local magnetic field allows a very large number of energetic electron, gyrosynchrotron emission can still be possible. Other physical parameters such as spectral index also need to be inspected in near future to confirm this.

Fine structures with even higher $T_\mathrm{B}$ ($\sim10^{13}$ K) than the background type IV may be generated by coherent ECM emission. This speculation needs some justification in the low-frequency regime because the ECM requires a relatively high ambient magnetic field common in the lower corona. \cite{2016A&A...589L...8M} used a PFSS model and calculated the ECM condition to be $\sim$ 500 MHz. However, \cite{2015A&A...581A...9R} used NLFF models and demonstrated that ECM condition can still satisfy at higher corona (1.2 Rs), which supports our conclusion. Another possible interpretation of the results is that a faint gyrosynchrotron background is superposed by high $T_\mathrm{B}$ plasma fine structures. However, we were not able to isolate this faint $T_\mathrm{B}$ background in the spectrum due to numerous fine structures superimposed on the continuum

We conclude that the background type IV is most likely generated by gyrosynchrotron emission, while the fine structures may still be generated by coherent ECM emission. However, future observations will be necessary to confirm our conclusion on the emission mechanism of moving type IV radio bursts and fine structure therein. { For example, statistical studies of a number of similar type IV bursts  will certainly contribute to our understanding of emission mechanism \citep[][]{salas2020polarisation, Kumari2021}}.

\begin{acknowledgements}
	This research received fund from UST overseas training program (South Korea), the Netherlands institute for Radio Astronomy (ASTRON), and Korea Astronomy and Space Science Institute under the R\&D program Development of a Solar Coronagraph on International Space Station (Project No. 2020-1-850-07) supervised by the Ministry of Science, ICT and Future Planning (South Korea). J.M. acknowledges funding by the BRAIN-be (Belgian Research Action through Interdisciplinary Networks) project CCSOM (Constraining CMEs and Shocks by Observations and Modeling throughout the inner heliosphere). A.K. acknowledges the European Research Council (ERC) under the European Union’s Horizon 2020 Research and Innovation Programme Project SolMAG 724391.
	We would like to thank SDO, LASCO, NRH and LOFAR operational team for providing open-access data, and SunPy team for the codes. In addition, we acknowledge Eduard Kontar and Nicolina Chrysaphi from University of Glasgow, and Jaeok Lee from KASI for their constructive suggestions. We also thank Sarrvesh Sridhar, Maaijke Mevius and Michiel Brentjens at ASTRON, Netherlands for their beneficial discussions and support. LOFAR, the Low Frequency Array designed and constructed by ASTRON, that has facilities in several countries, that are owned by various parties (each with their own funding sources), and that are collectively operated by the International LOFAR Telescope (ILT) foundation under a joint scientific policy.
\end{acknowledgements}

\bibliographystyle{spr-mp-sola}
\bibliography{reference}

\end{article}
\end{document}